\documentclass[a4paper,12pt]{article}

\usepackage[a4paper]{geometry}
\usepackage[dvipsnames]{xcolor}
\usepackage[utf8]{inputenc} 
\usepackage[hidelinks]{hyperref}
\usepackage{float}
\usepackage{amssymb,amsmath,amsthm}
\usepackage{amsfonts}

\usepackage[draft,colorinlistoftodos,shadow, textwidth=2.5cm, textsize=scriptsize]{todonotes}

\pdfoutput=1
\usepackage{graphicx}
\usepackage{graphicx, float, tabularx, booktabs}
\usepackage{color}
\usepackage{enumerate}
\usepackage[american]{babel}
\usepackage{stackrel}

\usepackage[numbers, compress]{natbib}
\usepackage{natbib}
\newcommand{\beq}{\begin{equation}}
\newcommand{\eeq}{\end{equation}}
\newcommand{\bea}{\begin{eqnarray}}
\newcommand{\eea}{\end{eqnarray}}
\newcommand{\be}{\begin{equation}}
\newcommand{\ee}{\end{equation}}
\newcommand{\ba}{\begin{eqnarray}}
\newcommand{\ea}{\end{eqnarray}}

\newcommand{\Mpl}{M_{\rm P}}

\newcommand{\ellp}{\ell_{\rm P}}

\usepackage{slashed}

\newcommand{\RS}{R_{\rm S}}

\usepackage{amsfonts}
\usepackage{epsfig}

\usepackage{graphicx, tabularx}
\vspace{-0.1cm}
\renewcommand{\thefootnote}{\fnsymbol{footnote}}

\begin{document}

\begin{center}
{\Large \textbf{
Combining the Generalized and Extended Uncertainty Principles 
}}
\end{center}

\vspace{-0.1cm}

\begin{center}
Bernard Carr$^{a}$\footnote{%
E-mail: \texttt{b.j.carr@qmul.ac.uk} } and  Jonas Mureika$^{b}$\footnote{%
E-mail: \texttt{jmureika@lmu.edu} }

\vspace{.6truecm}

\emph{\small  $^a$School of Physics and Astronomy, 
 Queen Mary University of London,\\[-0.5ex]
\small  Mile End Road, London E1 4NS, UK}\\[1ex]

\emph{\small  $^b$Department of Physics, Loyola Marymount University,\\[-0.5ex] Los Angeles,  California, USA}\\[1ex]



\end{center}

\begin{abstract}
\noindent{\small } \noindent
The Generalized Uncertainty Principle (GUP) and Extended Uncertainty Principle (EUP) are modifications to the Heisenberg Uncertainly Principle (HUP), expected to apply as the energy 
approaches the Planck scale.  Here we consider a possible combination of these modifications (GEUP) and analyse the implications in various regions of the $(\Delta x, \Delta p)$ plane. We also consider an alternative combination (EGUP) which exhibits duality between $\Delta p$ and $\Delta x$, showing that this has some unusual features. The parameters which describe these models are usually assumed to be positive but we extend our analysis to include negative values.  All these proposals entail a link between black holes and the various types of Uncertainty Principle.  In particular, the GEUP predicts a new kind of strong-gravity black hole and this implies an interesting  link between  black holes and elementary particles. 
\end{abstract}

\renewcommand{\thefootnote}{\arabic{footnote}} \setcounter{footnote}{0}

\section{Introduction}

The Heisenberg Uncertainty Principle (HUP) relates the uncertainty in the position of a particle $\Delta x$ to the uncertainty in its momentum $\Delta p$ and corresponds to $\Delta x \geq \hbar/ (2 \Delta p)$ where  $\hbar \equiv h/(2\pi)$ is the reduced Planck constant.  It can be understood heuristically as arising because the momentum of the photon used to determine the particle's position necessarily perturbs its momentum; it also derives from the commutation relation between the position and momentum operators $[\hat{x},\hat{p}] = i \hbar$.  The HUP leads to an expression for the reduced Compton wavelength $\hbar /(Mc)$ for a particle of rest mass $M$, this representing the scale over which its quantum wave function is smeared out.  This length scale
also arises in the Schr\"odinger and Klein-Gordon equations.  On the other hand, the (unreduced) Compton wavelength, $h /(Mc)$, arises in the scattering of photons off electrons and in the pair-production of electron-positron pairs.

\begin{figure}[h]
\includegraphics[scale=0.26]{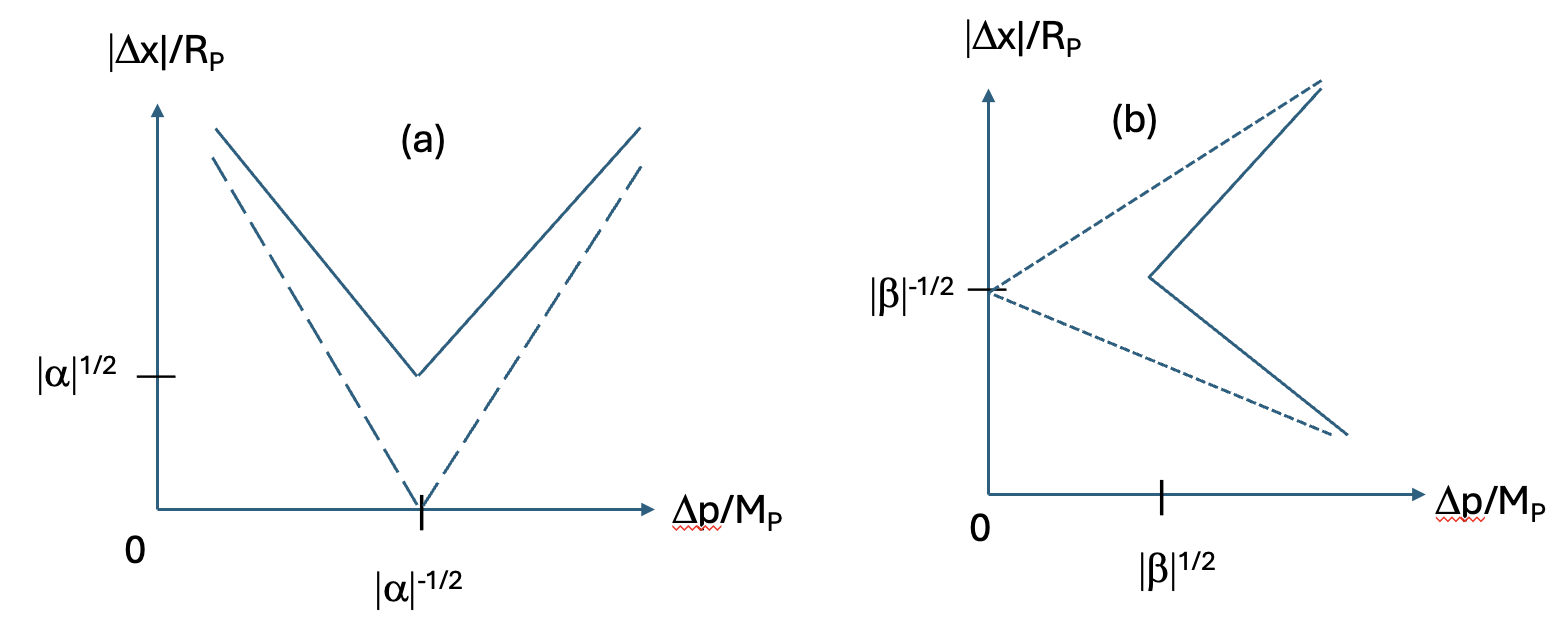}
\caption{Form of the function $\Delta x (\Delta p)$ for (a) GUP and (b) EUP.
 The modulus signs are necessary if $\alpha$ or $\beta$ is negative. The broken curves apply for negative $\alpha$ and the dotted curves for negative $\beta$.}
 \label{GUP/EUP}       
\end{figure}

The HUP must fail as one approaches
 the Planck scale ($\Mpl = \sqrt{\hbar c/G} \sim 10^{19}$~GeV, 
$\ellp = \sqrt{G \hbar/c^3} \sim 10^{-33}$~cm), since quantum gravity effects become important.  One possible modification is the Generalized Uncertainty Principle (GUP) and this corresponds to 
\beq
\Delta x \Delta p \geq \frac{\hbar}{2}\left(1+ \frac{\alpha \Delta p^2}{c^2 \Mpl^2}  \right) \, ,
\label{GUP1}
\eeq
which typically is written as
\beq
\Delta x \geq \frac{\hbar}{2}\left(\frac{1}{\Delta p} + \frac{\alpha \Delta p}{c^2 \Mpl^2}  \right) 
\label{GUP}
\eeq
for some dimensionless constant $\alpha$.  The second term can be understood heuristically as arising because the gravitational field of the probing photon adds an extra perturbation to the position of the particle.  It becomes important when the momentum
approaches $\Mpl \, c/\sqrt{ \alpha}$,  with $\Delta x$ reaching a minimum of $\sqrt{ \alpha} \, \ellp$ at this value. 
The GUP also corresponds to a modified commutation relation \cite{KMM95}
\beq
[\hat{x},\hat{p}] = i\hbar(\mathbb{I}+\alpha \ell_P^2 \hat{p}^2/ \hbar^2) \, .
\eeq
Indeed, for any operators $\hat{A}$ and $\hat{B}$,  an uncertainty relation can obtained from the expectation value of the commutator:
\beq
\Delta A \,  \Delta B \geq \frac{1}{2} \left|\langle[\hat{A},\hat{B}]\rangle\right|~.
\eeq
 The form of the function $\Delta x (\Delta p)$ in this case is shown in Fig.~\ref{GUP/EUP}(a). 
The parameter $\alpha$ could in principle be negative but in this case $\Delta x \rightarrow 0$ as $\Delta p \rightarrow  \Mpl \, c/\sqrt{|\alpha|}$ from below (left broken line).  Larger values of $\Delta p$ would lead to negative values of $\Delta x$, which is presumably unphysical.  However,  if one replaces $\Delta x$ by $|\Delta x|$ for larger values of $\Delta p$, the curve in Fig.~\ref{GUP/EUP}(a) extends into the  $\Delta p >  \Mpl \, c/\sqrt{|\alpha|}$ regime (right broken line). 
Since $\Delta x$ falls below the Planck length in this case,  quantum gravity effects may invalidate the analysis physically but it is still  illuminating mathematically.

If one extends the GUP expression to super-Planckian energies, it is natural to associate $\Delta x$ with the size of a black hole and this leads to the Black Hole Uncertainty Principle (BHUP) or Compton-Schwarzschild (CS) correspondence.  
More precisely, one identifies $\Delta x$ with the Schwarzschild radius and $\Delta p$ with $Mc$ where $M$ is the black hole mass.  However, this gives a radius which is 
half the Compton wavelength for $M \ll M_{\rm P}$ and 
$\alpha/2$ times the Schwarzschild radius for $M \ll M_{\rm P}$.  One can therefore reformulate the GUP,  so that it takes the form
\be
\Delta x \geq \hbar \left(\frac{\beta'}{\Delta p} + \frac{2 \Delta p}{c^2 \Mpl^2}  \right) 
\label{modGUP}
\ee
(i.e.  one associates a free parameter $\beta '$ with the first term). The justification for this is that the Schwarzschild radius is known exactly, whereas the Compton wavelength arises in different contexts, so its definition is less precise.  However, for the purposes of this paper we will adopt Eq.~\eqref{GUP} rather than Eq.~\eqref{modGUP}.

Another possible modification is the  Extended Uncertainty Principle (EUP),
\beq
\Delta x \Delta p \geq \frac{\hbar}{2}\left(1 +  \frac{\beta \Delta x^2}{ \ellp^2} \right) \, ,
\label{EUP1}
\eeq
which can also be expressed 
as
\beq
\Delta x \geq \frac{\hbar}{2}\left(\frac{1}{\Delta p}  +  \frac{\beta \Delta x^2}{ \ellp^2 \Delta p} \right) \ 
\label{EUP}
\eeq
for some dimensionless constant $\beta$. The second term becomes important as $\Delta x$ approaches $L_* \equiv \ellp/\sqrt{\beta}$, with $\Delta p$ having a minimum of   $\sqrt{\beta} \Mpl  c $ at this value.  
For $\Delta x \gg L_*$, one has $\Delta x <  2 G \Delta p/(\beta c^3)$, which resembles the GUP relation $\Delta x > G \alpha \Delta p/(2c^3)$ but with a reversal of the inequality sign.  So while the GUP introduces a minimum length, the EUP introduces a minimum momentum, as illustrated in Fig.~\ref{GUP/EUP}(b). One can regard Eq.~\eqref{EUP}, as a quadratic equation for $\Delta x$ in terms of $\Delta p$, in which case the two solutions correspond to the asymptotic forms $\Delta x \propto \Delta p^{-1}$ and $\Delta x \propto \Delta p$ for positive $\beta$.  If the parameter $\beta$ is negative, 
$\Delta x > 0$ only for the $\Delta x \propto \Delta p^{-1}$ solution and $\Delta p \rightarrow 0$ as $\Delta x \rightarrow \ellp/ \sqrt{|\beta|}$ (bottom dotted line). However, one retains the $\Delta x \propto \Delta p$ solution if one uses $|\Delta x|$ (top dotted line). 

The EUP corresponds to a commutation relation of the form
\beq
[\hat{x},\hat{p}] = i \hbar \left[\mathbb{I}+ \frac{\beta}{L_*^2} \hat{x}^2\right] \, .
\eeq
Although there is no heuristic rationale for this form, 
it may be seen as an algebraic completion
of the GUP relation that produces a ``symmetry'' in minimum values.  In the usual form of the EUP, the constant $\beta$ is very small.  This is because the Compton expression must apply for ordinary particles, which have much less than the Planck mass, and this requires $\beta \ll 1$.   There is also an associated length-scale $L_*$, which may be very large and even astronomical:
\beq
\beta \equiv (\ellp/ L_*)^{2} \sim  10^{-114} (L_*/ \rm{Mpc})^{-2} \, .
\label{beta}
\eeq
Indeed,  the value of $L_*$ may be set by the cosmological constant $\Lambda$, as outlined in Ref.~\cite{Pantig:2024asu}, through a relation dubbed the Dabrowski-Wagner-Mureika formalism:
\beq
L_* = \sqrt{\frac{6\pi^2\beta}{\Lambda}} \quad \Rightarrow \quad \beta = \left(\frac{\Lambda}{6 \pi^2} \right)^{1/2}  \ellp \sim 10^{-120} \,, 
\label{dwm}
\eeq
where the second relationship follows from Eq. ~\eqref{beta} and assumes the Universe is dominated by its vacuum energy.
However,  in principle $L_*$ could be a particle scale, in which case $\beta$ would be much larger.  For example, the Compton expression extends down to the electron mass for $\beta < 10^{-44}$ and this implies that $L_*$ must exceed the Compton wavelength of the electron. 

As in the GUP case,  it is natural to extend the EUP to super-Planckian scales,  thereby providing a link between black holes and elementary particles.
 However,  this is not be possible with the standard identification of $\Delta x$ and $\Delta p$  with the black hole radius and mass.  
On the one hand, if $\beta \sim 1$, the Compton wavelength is only defined in the super-Planckian regime.  On the other hand, if $\beta \ll 1$,  the  black hole radius is much larger than Schwarzschild value.  
This paper provides a possible resolution of this problem,  although we will find that this has some unusual  consequences.

There is, however, another version of the EUP, based on 
 the corpuscular Bose-Einstein condensate model of black holes \cite{Dvali:2011aa}, which gives a very different relation between the black hole radius and mass, merely yielding a modification to the Schwarzschild metric for large masses \cite{Mureika_2019}. 
In this case,  the weak field limit of the EUP-Schwarzschild metric 
may offer an alternative to dark matter, with the length scale $L_*$ being used to determine the radius at which galactic rotation curves plateau~\cite{Mureika_2019}.   
Subsequent studies based on this finding have 
included analyses of the Mercury's perihelion shift, the Shapiro time delay, and the precession of the S2 star around Sgr A* \cite{Okcu:2022sio}. The authors of Ref.~\cite{luxie} considered EUP-gravity effects on the strong lensing of Sgr A* and M87. Reference \cite{Illuminati:2021wfq} explored lensing effects, as well as binary pulsar motion and solar-spin precession. 
All these references place {\it lower} bounds on the EUP length scale $L_*$. 

In fact, as discussed in the Appendix,  the corpuscular theory can be used to justify either version of the EUP, 
or indeed the GUP,  so it does not predict the metric unambiguously.  Although we do not focus on the metric of Ref.~\cite{Mureika_2019} in this paper,  we discuss it further in the  Appendix.
Note that both the GUP and EUP of Ref.~\cite{Mureika_2019} can be interpreted as resulting from non-local gravitational effects in black holes \cite{Capozziello:2025iwn}. Corrections to the static Newtonian potential were obtained in each case, and constraints on corresponding black hole horizons were derived. In particular, this work showed that 
the values of the GUP and EUP parameters 
can be approximated as
\be
\alpha\approx-\frac{M^2}{M_\textup{P}^2}, \quad
\beta \approx  -0.2\,\frac{L_*^2c^4}{G^2M^2}\approx-137 \, \frac{M_\textup{P}^2}{M^2}\frac{L_*^2}{\ellp^2} \, .
\label{eupapprox}
\ee
If Eq.~\eqref{beta} applies, $L_* \sim (m/M_{\rm P})^{1/2} \ellp$.  
Both expressions set the parameters in terms of fundamental constants associated with each framework. 
These relations are of particular interest because they suggests that both $\alpha$ and $\beta$ are negative, whereas traditionally they are assumed to be positive. Such a proposal was also raised in Ref.~\cite{Fragomeno:2024tlh} with respect to the GUP parameter. 

It is clearly interesting to combine the GUP and EUP effects (i.e.  to include both $\alpha$ and $\beta$). This is termed the Generalized Extended Uncertainty Principle (GEUP). One motivation for this is that it restores the link between black holes and elementary particles, which we have seen does not apply for EUP alone 
Some literature has already been devoted to the GEUP. 
For example,  Ref.~\cite{Lobos:2022jsz} has addressed the impact of the GEUP on black hole shadow formation in the micro- and macro-lensing regimes.  In Ref.~\cite{Pachol:2024hiz}, the link between the GEUP,  Liouville theorems, and quantum density of states is explored. 
In Ref.~\cite{Casadio:2025sjp} constraints are placed on the compactness of astrophysical objects through the known limits on the GUP and EUP parameters.  

In Section 2 we analyse the form of the $\Delta x(\Delta p)$ relationship in the GEUP case,  going beyond previous works by studying the quantum black hole regime.
We also allow for negative values of $\alpha$ and $\beta$, which has not been considered before but is not physically excluded.  
There is another combination, which we term the Extended Generalised Uncertainty Principle (EGUP).  This is not so well motivated physically but it has interesting symmetry properties, so we analyse this in Section 3, again allowing for negative values of $\alpha$ and $\beta$.  
In Section 4,  we examine the implications of the GEUP and EGUP for black holes,  thereby the extending the BHUP correspondence  beyond the GUP case.  In Section 5, we discuss the implications of this for the Hawking temperature,  again extending the analysis beyond the GUP case.  We draw some general conclusions in Section 6.

\section{Generalized-Extended Uncertainty Principle}

A modification of the HUP which combines the GUP and EUP is \cite{KMM95} 
\beq
\Delta x \Delta p \geq \frac{\hbar}{2} \left(1+ \frac{\alpha \Delta p^2}{c^2 \Mpl^2} + \frac{\beta \Delta x^2}{\ellp^2}  + \gamma \right) \, ,
\label{GEUPX}
\eeq
where $\gamma$ is an overall constant that may be some function of $x$ and/or $p$.  This 

may be derived from the uncertainty relation~\cite{KMM95}
\beq
[\hat{x},\hat{p}] = i\hbar\left(\mathbb{I} + \frac{ \alpha \ell_P^2 \hat{p}^2}{\hbar^2} + \frac{\beta}{\ellp^2}\hat{x}^2\right)~
\eeq
provided 
\beq
\gamma = \alpha \ell_P^2 \langle p\rangle^2/ \hbar^2 + \beta \langle x\rangle^2/\ellp^2 \, .
\label{gammaeq}
\eeq
Note that $\gamma = 0$ for $\langle p\rangle = \langle x\rangle =0$, implying a wavefunction which is antisymmetric about the origin.  Alternatively,  $\gamma = 0$ may be ensured if
$\alpha$ and $\beta$ have opposite signs. Indeed, 
some works have suggested that either or both of these parameters may be negative.
In this case, 
 Eq. (\ref{gammaeq}) implies
\beq
\alpha = -\frac{\beta\hbar^2}{\ell_P^4}\frac{\langle x \rangle^2}{\langle p \rangle^2} \, .
\eeq

\begin{figure}[h]
\includegraphics[scale=.3]{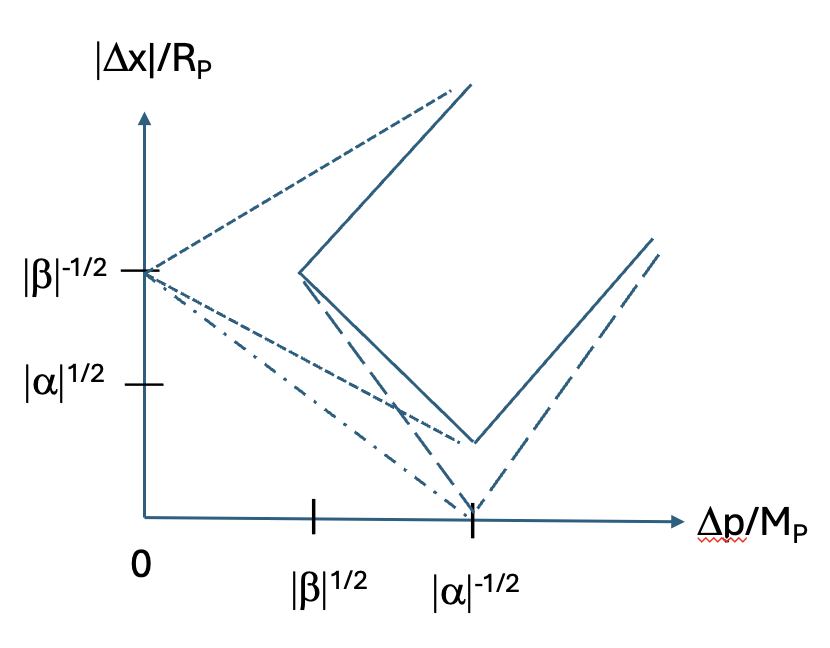}
\caption{Qualitative form of the function $\Delta x (\Delta p)$ in the GEUP case for $\alpha >0$ and $\beta >0$ (solid lines), 
$\alpha < 0$ and $\beta >0$ (broken lines), $\alpha >0$ and $\beta < 0$ (dotted  lines), $\alpha <0 $ and $\beta < 0$ (upper dotted  plus broken-dotted plus right broken lines).  The modulus signs are necessary in all but the first case.}
 \label{EUP/GUP}       
\end{figure}

It is the simplified version of Eq.~(\ref{GEUPX}) with $\gamma =0$ which has been dubbed the Generalized Extended Uncertainty Principle (GEUP).  In this case
\beq
\Delta x \Delta p \geq \frac{\hbar}{2} \left(1 + \frac{\alpha \ell_P^2}{\hbar^2} \Delta p^2 + \frac{\beta}{\ell_P^2}\Delta x^2 \right) \, ,
\label{geup}
\eeq
so $\beta =0$ corresponds to the GUP and  $\alpha =0$ to the EUP. 
One might anticipate the qualitative form of the function $\Delta x(\Delta p)$ in this case by combining the qualitative forms indicated in Fig.~\ref{GUP/EUP}. 
For different combinations of signs for $\alpha$ and $\beta$, 
this gives the various forms indicated  in Fig.~\ref{EUP/GUP}, with a minimum in both $\Delta x$ and $\Delta p$.

We now present a more precise analysis.
Rearranging Eq.~(\ref{geup}) and considering the equality limit gives 
\beq
\Delta x = \frac{\hbar}{2}\left(\frac{1}{\Delta p} + \frac{\alpha \Delta p}{c^2 \Mpl^2} +  \frac{\beta \Delta x^2}{ \ellp^2 \Delta p} \right) \ .
\label{quadratic}
\eeq
Regarding this as a quadratic equation for $\Delta x$ in terms of $\Delta p$ leads to 
\beq
 \Delta x = \frac{G}{\beta c^3} \left( \Delta p \pm \sqrt{(1- \alpha \beta) \Delta p^2 - \beta c^2 \Mpl^2} \right) \, .
\label{quadx}
\eeq
For positive $\alpha$ and $\beta$, this shows that there is only a real solution for $\alpha \beta < 1$ and
\beq
\Delta p < \sqrt{\frac{\beta}{1-\alpha \beta}} \, \,  \Mpl   c \, .
\eeq 
In the usual form of the EUP the constant $\beta$ is very small, which ensures that the condition $\alpha \beta < 1$ is likely to be satisfied.  

\begin{figure}[h]
\includegraphics[scale=.4]{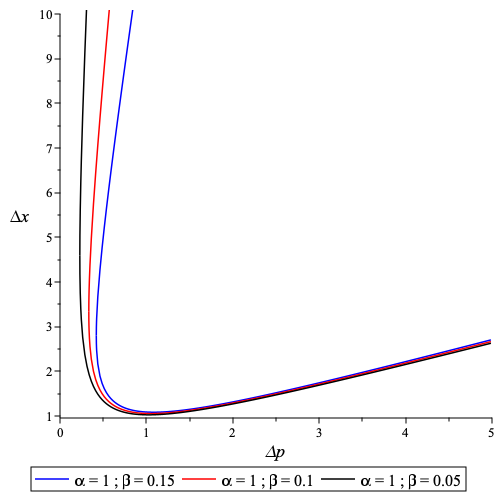}
\includegraphics[scale=.4]{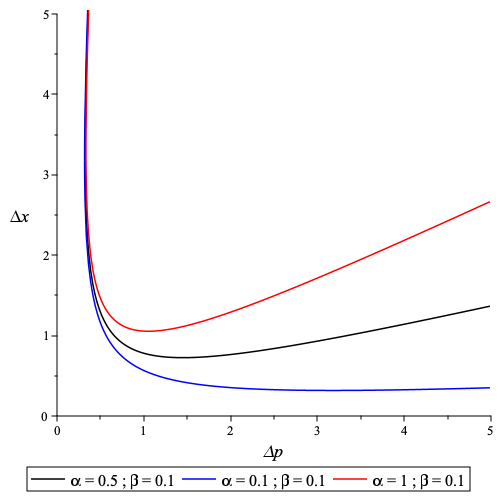}
\caption{Form of the function $\Delta x (\Delta p)$ for GEUP with $\alpha = 1$ and different values of 
$\beta$ (left) and $\beta= 0.1$ and different values of 
$\alpha$ (right).  Note that $\alpha\beta < 1$ is satisfied for all cases. }
 \label{GEUP}       
\end{figure}

The form of the function $\Delta x(\Delta p)$ is indicated in Fig.~\ref{GEUP} for various positive values of $\alpha$ and $\beta$ with $\alpha \beta < 1$.  
In order to explain these forms,  we note  the gradient is
\beq
\frac{\Delta x}{\Delta p} = \frac{G}{\beta c^3}\left[1\pm \sqrt{1-\beta \left( \alpha + \frac{\Mpl^2 c^2}{\Delta p^2} \right)} \, \right] \, .
\eeq
Since we are interested in the large mass behaviour of the GEUP ({\it i.e.} $\Delta p \rightarrow \infty$) and assume $\alpha \beta \ll 1$,  this can be approximated as
\beq
 \frac{\Delta x}{\Delta p} \approx \frac{G}{\beta c^3} \left[ 1\pm \left(1- \frac{\beta \alpha}{2} - \frac{\beta \Mpl^2c^2}{2 \Delta p^2} \right) \right]
 \approx  
\begin{cases} 
2G/(\beta c^3) & (+) \\ \hbar/ (2 \Delta p^2) & (-) \\ \alpha G/(2 c^3) & (-) 
 \, ,
\end{cases}
\label{geupgradcases}
\eeq
where the transition between  the first two cases is at $\Delta p \approx c \sqrt{\beta} \Mpl$ and the transition between the last two cases is at $\Delta p \approx c \Mpl/ \sqrt{\alpha}$ (cf.  Fig.~\ref{GUP/EUP}). Therefore the curves in Fig.~\ref{GEUP} have three asympotic slopes: $\alpha/2$ in the far-right region,  $2/\beta$ in the top-left region and $-1$ in the lower-left region.  Note that the condition $\alpha \beta < 1$ ensures that the first slope is always less than the second, so the two asymptotic curves never intersect.  

Of particular interest are the upper ``branches'' of the GEUP curves in Fig.~\ref{GEUP}. In the standard GUP, the $\Delta x(\Delta p)$ curve represents the large-scale Schwarzschild behavior for black holes and the small-scale Compton
behavior for particles.  Reference~\cite{CMN15} presented a metric solution known as the `M+1/M' model in which the small-mass limit was proposed to represent sub-Planckian black hole solutions. In Fig.~\ref{GEUP}, however, the curves do not asymptote to the $\Delta x$ axis, as in the GUP case, but 
reach a minimum $\Delta p$ ({\it i.e. } a minimum black hole mass) and then
curve back to increasing $\Delta p$. We propose that these correspond to black hole solutions in a strong gravity regime. 

\begin{figure}[h]
(a)\includegraphics[scale=.35]{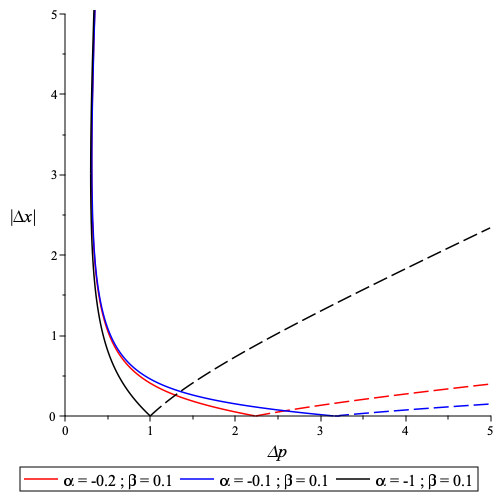}
(b)\includegraphics[scale=.35]{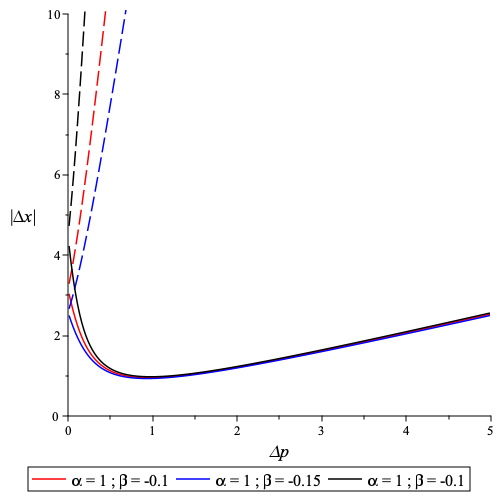}
(c)\includegraphics[scale=.35]{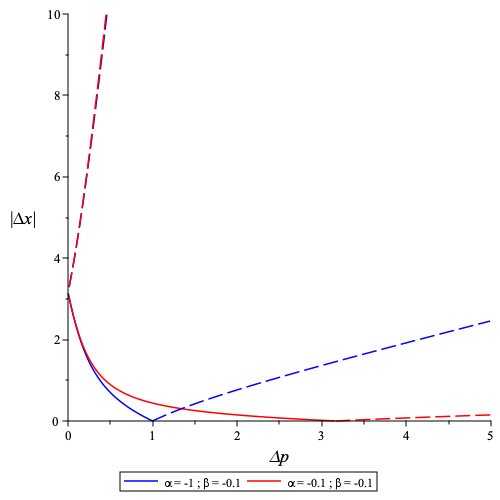}
\caption{GEUP curves for (a) $\alpha <0$ and $\beta>0$, (b) $\alpha >0$ and $\beta<0$ and (c) $\alpha <0$ and $\beta<0$.  The constraint $|\alpha\beta|<1$ is maintained and $|\Delta x|$ is required if $\alpha$ and/or $\beta$ is negative (dashed lines). }
 \label{GEUPnegab}       
\end{figure}

Since it has been shown that the parameters $\alpha$ and $\beta$ may be negative, we also present plots for this case in Fig.~\ref{GEUPnegab}. From Eq.~\eqref{quadx},  if either $\alpha$ or $\beta$ is negative (but not both), the constraint $\alpha\beta < 1$ is removed.  If only $\alpha < 0$,  
 the GEUP curves must satisfy the constraint 
\beq
\Delta p > \sqrt{ \frac{\beta}{1+|\alpha| \beta}} \, \, c M_{\rm P}
\eeq
for $\Delta x$ to be real. In this case,  $\Delta x$ has a minimum at zero and the solution cannot be extended beyond that point,  as indicated by the solid  lines Fig.~\ref{GEUPnegab}(a).  However, if one uses $| \Delta x|$,  the solution can be extended into the super-Planckian regime,  as indicated by the broken curves.  These forms correspond to the broken curves anticipated in Fig.~\ref{GUP/EUP}(a).  Although Eq.~(\ref{geupgradcases}) implies that the third gradient would be negative ({\it i.e.} $\Delta x / \Delta p <0$), this part of the curve is excluded if one uses $\Delta x$ and the gradient is positive if one uses $|\Delta x|$.  

If only $\beta < 0$, there are no constraints on $\Delta p$, so the whole range of values is covered, as indicated by  Fig.~\ref{GEUPnegab}(b).
On the other hand,  only the negative sign in Eq.~\eqref{quadx} allows $|\Delta x| >0$ and Eq.~(\ref{geupgradcases}) then implies that 
 the gradient is necessarily negative in the small $\Delta p$ regime.  It therefore never changes sign, as in the $\beta >0$ case.  
However, the positive sign in Eq.~\eqref{quadx} does give a possible solution if one uses $|\Delta x|$. In this case, the positive gradient at small $\Delta p$ is allowed, corresponding to the dotted curves in Fig.~\ref{GEUPnegab}(b).
 If both $\alpha$ and $\beta$ are negative, one finds the form shown in Fig.~\ref{GEUPnegab}(c). 
The qualitative form of $\Delta x(\Delta p)$ for every combination of signs for $\alpha$ and $\beta$ can be inferred from Fig.~\ref{GUP/EUP} and is indicated more precisely in Fig.~\ref{GEUPnegab}.

\section{Extended Generalized Uncertainty Principle}
\label{Sec:EGUP}

\begin{figure}[h]
(a) \includegraphics[scale=.4]{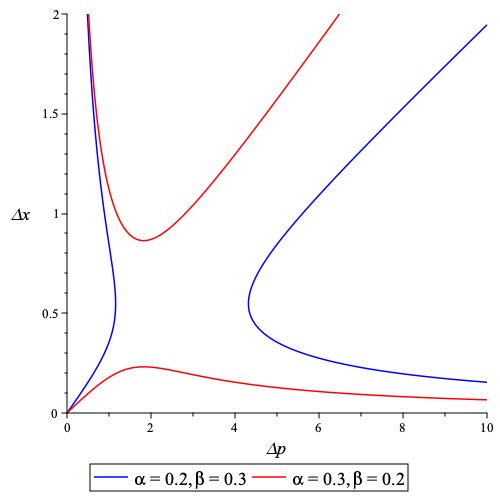}
(b) \includegraphics[scale=.4]{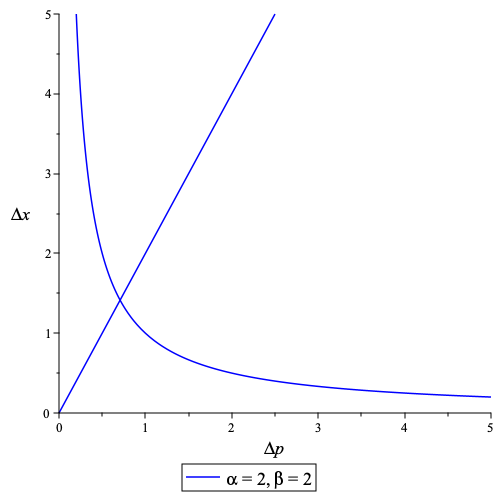}
\caption{Form of the functions $\Delta x (\Delta p)$  in EGUP case for  (a) $\alpha > \beta$ and $\alpha < \beta$  and (b) $\alpha = \beta$.  }
 \label{EGUP}       
\end{figure}

One could envisage another version of the Uncertainty Principle as 
\beq
\Delta x + \frac{\beta \ellp^2}{\Delta x} \geq \frac{\hbar}{2 \Delta p} + \frac{\alpha \Delta p \, \ellp^2}{2 \hbar} \, .
\eeq
We term this the Extended General Uncertainty Principle (EGUP) and its physical motivation is that it exhibits duality between $\Delta x$ and $\Delta p$. It reduces to the GUP when $\beta =0$ but is not identical to the EUP when $\alpha =0$.  
Solving for $\Delta x$, we find
\be
\Delta x = \frac{1}{4} \left(\frac{\hbar}{\Delta p}+\frac{\alpha\Delta p \ell_P^2}{\hbar}\right)\left(1\pm\sqrt{1-\frac{8\beta\ell_P^2}{\left(\frac{\hbar}{\Delta p}+\frac{\alpha\Delta p \ell_P^2}{\hbar}\right)^2}}\right) \, .
\label{EGUPeqn}
\ee
In this case, the function $\Delta x(\Delta p)$ has the form shown in Fig.~\ref{EGUP}. 
The form of the solutions depends on whether $\alpha$ is greater or less than $\beta$. For $\alpha > \beta$, the particle and black hole regions are connected but there is also a sub-Planckian domain, which is presumably unphysical.  This is shown by the red lines in Fig.~\ref{EGUP}(a).  For $\alpha < \beta$, the particle and black hole regions are disconnected, as shown by the blue lines in Fig.~\ref{EGUP}(a). This is similar to the behavior exhibited by the $M+1/M$ solutions
 for charged and rotating black holes~\cite{CMMN20}. The transition at $\alpha = \beta$  is shown in Fig. \ref{EGUP}(b). 

\begin{figure}[h]
(a) \includegraphics[scale=.35]{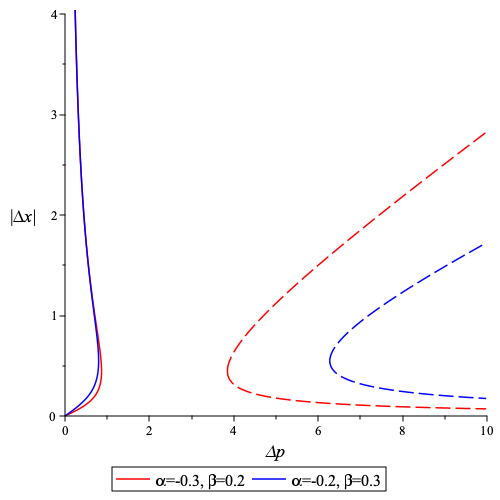}
(b) \includegraphics[scale=.35]{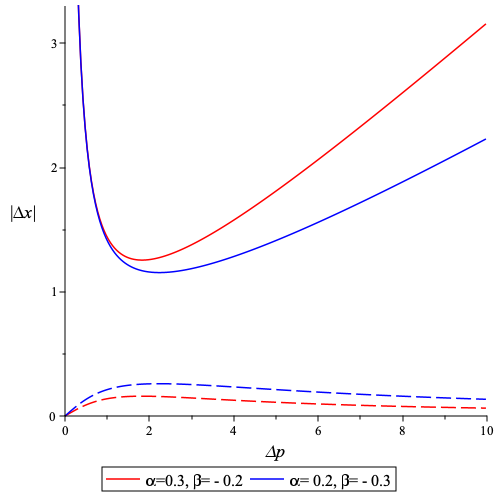} \\
(c) \includegraphics[scale=.35]{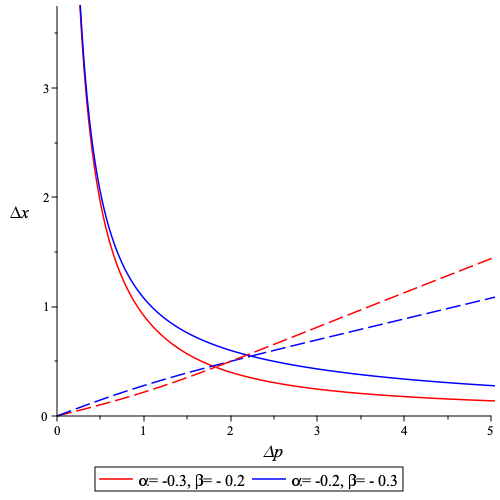}
\caption {EGUP parameter space for (a) $\alpha < 0,\beta >0$, (b) $\alpha>0,\beta<0$ and (c) $\alpha<0,\beta<0$.  The dashed curves apply if one uses $|\Delta x|$.}
 \label{EGUPnegab}       
\end{figure}

As with the GEUP case, we also consider 
negative values of $\alpha$ and $\beta$, as shown in Fig.~\ref{EGUPnegab}. 
If $\beta$ is negative and $\Delta x$ is required to be positive,  then only the negative sign in Eq.~\eqref{EGUPeqn} is allowed,  giving the upper red (GUP-type) curve in Fig.~\ref{EGUPnegab}(b). 
In this case, 
the slope of the curve approaches $\alpha$ for large $\Delta x$.
However,  using $|\Delta x|$ would also give solutions corresponding to the lower red curve in Fig.~\ref{EGUPnegab}(b). 
If $\alpha$ is negative and $\Delta x$ is required to be positive,  both the positive and negative sign in Eq.~\eqref{EGUPeqn} is allowed but there is a maximum value for $\Delta p$. This corresponds to the left blue curve in Fig.~\ref{EGUPnegab}(a).   
However,  using $|\Delta x|$ would also give the broken blue curve  in Fig.~\ref{EGUPnegab}(a).  If both $\alpha$ and $\beta$ are negative,  Fig.~\ref{EGUPnegab}(c) applies.

\section{Extending to black hole region}

Under the transformation $\Delta p \rightarrow c M$ and $\Delta x \rightarrow R$,  the function $\Delta x(\Delta p)$ also specifies the form of the function $R(M)$.  In the context of the GUP,  $R(M)$ approximates the Compton wavelength $R_{C}$ of a particle for $M < \Mpl$ and the Schwarzschild radius of a black hole $R_S$ for $M > \Mpl$. This suggests a unification of these two length scales, with a smooth minimum at around the Planck length.  
This is termed the Black Hole Uncertainty Principle (BHUP) or Compton-Schwarzschild (CS) correspondence.  As discussed in our previous work~\cite{CMN15,CMMN20,Carr:2024brs}, this suggests that the Compton wavelength can also be interpreted as a black hole radius in some sense,  implying some link between elementary particles and sub-Planckian black holes. 
In this section we extend this idea to the EUP, GEUP and EGUP.
In this and the following section, we assume that $\alpha$ and $\beta$ are positive.

The simplest way to realise the BHUP correspondence in the context of the GUP,  i.e.  to ensure a smooth minimum for $R$ in the $R(M)$ diagram,  is to invoke a metric
of the form
\beq
ds^2 = F(r)~dt^2 - \frac{dr^2}{F(r)}-r^2 d^2\Omega
\label{metric}
\eeq
where 
\beq
F(r) = 1-\frac{2GM}{c^2 r} \left(1 + \frac{\beta'}{2}\frac{\Mpl^2}{M^2}\right) \, .
\label{gupmetric}
\eeq
In this case,  the  horizon radius of the corresponding black hole is
\beq
R_H = \frac{2GM}{c^2} + \frac{\beta' \hbar}{Mc}
\eeq 
where $\beta'$ has a prime to distinguish it from $\beta$ in the EUP.  This encapsulates both the expected macroscopic (first term) and microscopic (second term) behavior of the horizon.  
For notational convenience, 
we here  associate the free parameter with the first term and call it $\alpha$, so that
\beq
R_H = \frac{\alpha GM}{2 c^2} + \frac{\hbar}{2Mc} \, .
\label{GUPradius}
\eeq 
Although this means that neither the Compton wavelength nor the Schwarzschild radius has the standard coefficient, this is the easiest way to ensure consistency with our earlier discussion. 

We now  extend this argument to the EUP but there is an important difference in this case. 
If we identify $\Delta x$ with $R_H$ and $\Delta p$ with $Mc$, then Eq.~\eqref{EUP} implies 
\beq
R_{H} = \frac{G}{\beta c^2} \left( M \pm \sqrt{M^2 - \beta \Mpl^2} \right)
\approx 
\begin{cases}
\hbar/(2Mc) & (-) \\
2GM/(\beta c^2) & (+) \, ,
\end{cases} 
\label{EUPradius}
\eeq
where we require $M > \sqrt{\beta}  M_{\rm P}$ and the last step applies for $M \gg \sqrt{\beta}  M_{\rm P}$.  Although this has similar asymptotic limits to the BHUP case, there are two crucial differences: $\alpha$ is replaced by $1/\beta$ and there are two values of $R_H$ for each value of $M$. This is illustrated by a comparison of Figs.~\ref{GUP/EUP}(a) and (b).
One still has a Compton and Schwarzschild line but they cannot represent a link between the usual type of particles and black holes.  
If one chooses 
$\beta =1$,  so as give the exact Schwarzschild expression,  the Compton expression does not extend below the Planck mass.   
On the other hand,  if $\beta \ll 1$,
the black holes described by Eq.~\eqref{EUPradius} are not of the usual kind because the effective gravitational constant is much larger than observed.  More precisely,  it becomes
\beq
G_* = G / \beta = G (L_*/\ellp)^{2} \, ,
\label{G*}
\eeq
where the length scale $L_*$ is defined by Eq.~\eqref{beta}. 
 If this is an astronomical scale, as assumed in Ref.~\cite{Mureika_2019},  then $\beta \ll 1$ and $G_* \gg G$, so these black holes have a tiny mass but a large radius:
\beq
R_S \sim G_* M/c^2 \sim (M/M_*) L_* \quad {\rm with} \quad  M_*  \sim 10^{-61} (L_*/ \rm{Mpc})^{-1} {\rm gm} \, .
\eeq
Therefore the EUP cannot be relevant to the black holes in our universe. Since  we know these exist, this means that one {\it requires} the GEUP, which we discuss below.

We note that Eq.~\eqref{EUPradius} is different from the $R_H(M)$ relation for EUP black holes formulated in Ref.~\cite{Mureika_2019}. This is predicated on the corpuscular model \cite{Dvali:2011aa} and discussed further in the Appendix. However,  that version of the EUP does not attempt to describe partices at all because the standard Schwarzschild expression extends down to arbitrarily small mass. 

In the GEUP case,  
identifying $\Delta x$ with $R_H$ and $\Delta p$ with $Mc$ in Eq.~\eqref{quadx}
gives
\beq
R_{H} = \frac{G}{\beta c^2} \left( M \pm \sqrt{(1- \alpha \beta) M^2 - \beta \Mpl^2} \right) \, .
\label{GEUPrad}
\eeq
In the limit $\alpha \rightarrow 0$,  this gives the EUP expression~\eqref{EUPradius}.  In the limit $\beta \rightarrow 0$, the positive sign gives a divergence and is inapplicable but the negative sign gives
\be
R_{H} = \frac{G}{2 c^2} ( \alpha M + M_{\rm P}^2/M) \approx
\begin{cases}
\alpha GM/(2c^2) &  (M \gg  M_{\rm P}/ \sqrt{\alpha}) \\
\hbar/(2Mc) & (M \gg  M_{\rm P}/ \sqrt{\alpha}) \, ,
\end{cases}
\ee
which is equivalent asymptotically to Eq.~\eqref{GUPradius}. 
More generally, for positive $\alpha$ and $\beta$, there is only a real solution for $\alpha \beta < 1$ and 
\beq
M > M_* \equiv  \sqrt{\frac{\beta}{1-\alpha \beta}} \, \Mpl \, .
\label{Mstar}
\eeq
In this case,  for $\alpha \beta \ll 1$, we have the following asymototic forms: 
\beq
R_{H} \approx 
 \begin{cases}
\alpha GM/(2c^2) & (-,  M  \gg \Mpl/ \sqrt{\alpha})\\
\hbar/(2Mc) & (-,  M  \ll \Mpl/ \sqrt{\alpha}) \\
2GM/(\beta c^2) & (+, M \gg M_*)
 \, .
\label{GEUPradius}
\end{cases}
\eeq
The form of the function $R_{H}(M)$ is similar to that of the function $\Delta x(\Delta p)$ shown in Fig.~\ref{GEUPnegab}, so this is  not plotted explicitly.
Apart from the numerical coefficients, the first expression in Eq.~\eqref{GEUPradius} is the usual black hole horizon radius and the second expression is the Compton wavelength. The Schwarzschild expression might suggest $\alpha = 4$ but this relates to the interpretation of the constants in Eq.~\eqref{GUP}. 
As explained earlier,  if the free constant is associated with $M^{-1}$ term, the coefficient of the $M$ term can assume the standard Schwarzschild value~\cite{Carr:2024brs}.  

In the EGUP case,  Eq.~\eqref{EGUPeqn} with $\Delta x \rightarrow R_H$ and $\Delta p \rightarrow Mc$ implies a black hole radius
\be
R_H = \frac{1}{4} \left(\frac{\hbar}{M c}+\frac{\alpha G M}{c^2}\right)\left(1\pm\sqrt{1-\frac{8\beta M^2}{\left(\Mpl +\alpha M^2/ \Mpl\right)^2}}\right) \, .
\label{EGUPradius}
\ee
The form of $R_H(M)$ for various values of $\alpha$ and $\beta$ can then be inferred from Figs. ~\ref{EGUP} and \ref{EGUPnegab}.  This is similar to the form of $R_H(M)$ for charged and rotating black holes in the `$M+1/M$' model \cite{CMMN20}.  

\section{Black hole temperature}

\begin{figure}[h]
\includegraphics[scale=.22]{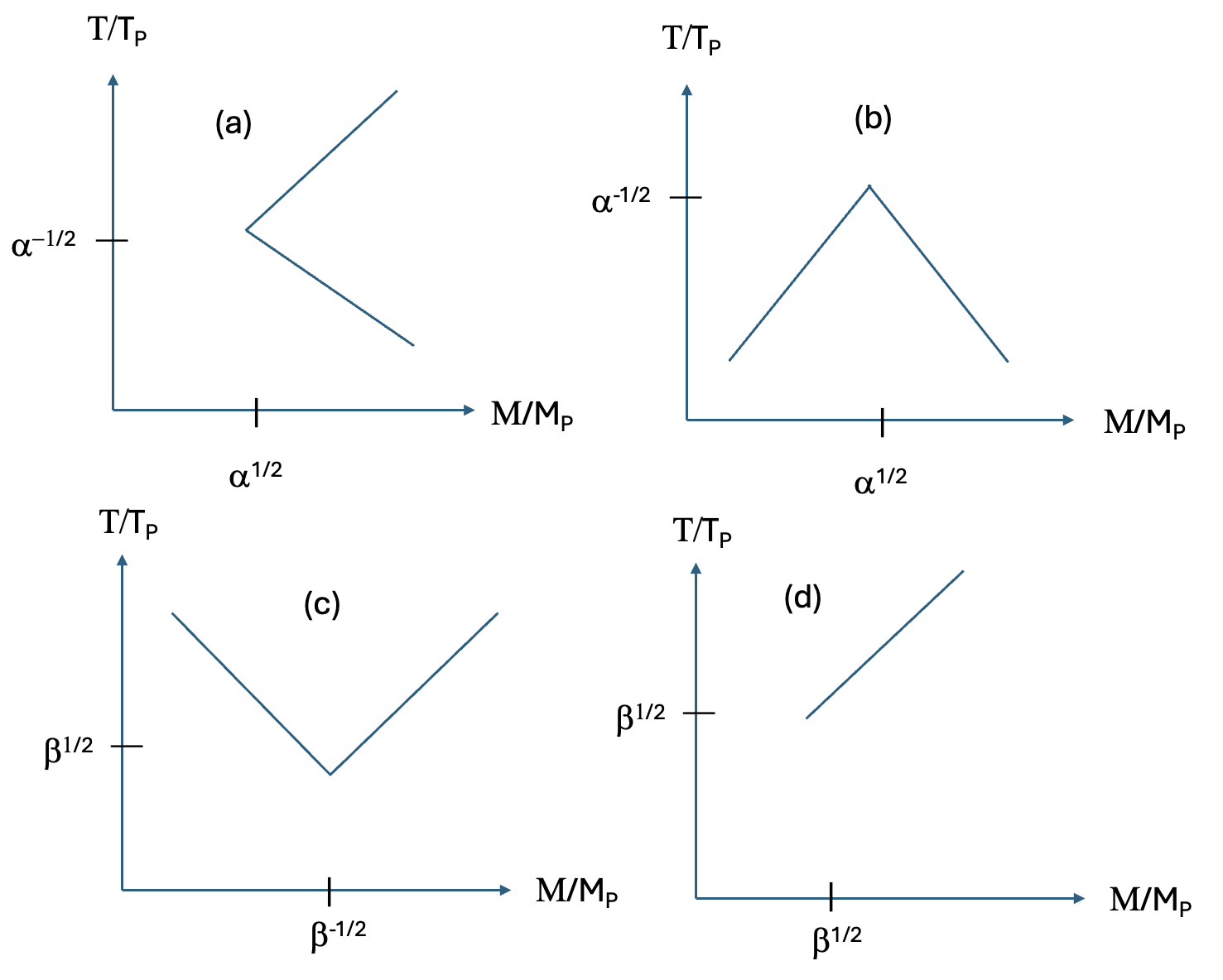}
\caption{The form of the temperature function $T(M)$ for: (a) GUP model with standard black hole radius; (b) BHUP model with modified black hole radius; (c) EUP model with standard black hole radius; (d) EUP model with modified black hole radius.  The parameters $\alpha$ and $\beta$ are assumed to be positive.}
 \label{flatGUP}       
\end{figure}

\begin{figure}[h]
(a) \includegraphics[scale=.29]{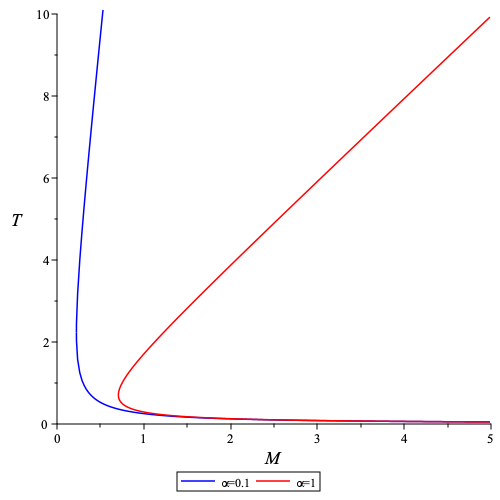}
(b) \includegraphics[scale=.29]{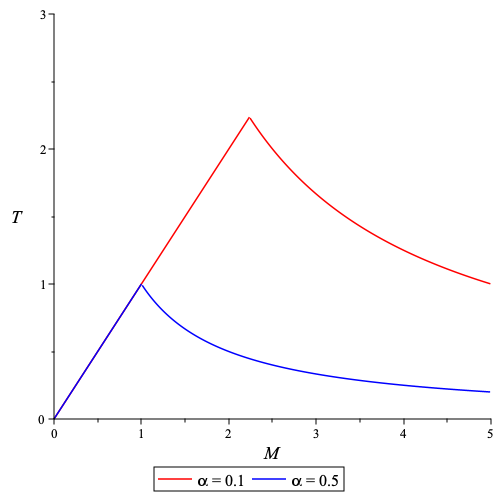} \\
(c) \includegraphics[scale=.29]{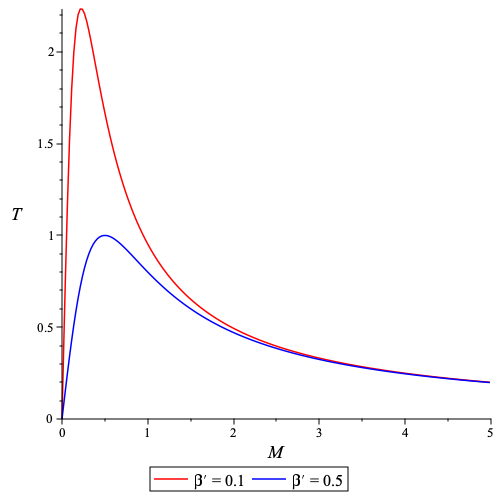}
(d) \includegraphics[scale=.29]{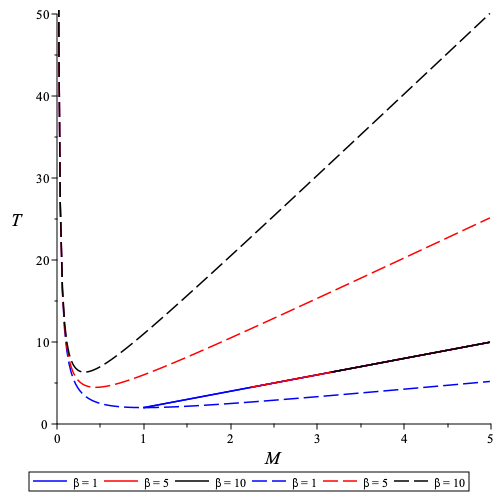}
\caption{Numerical calculations of the temperature function $T(M)$ for: (a) GUP model with standard black hole radius; (b) GUP model with modified black hole radius (i.e.  BHUP model) and Adler temperature; (c) BHUP model with modified black hole radius and surface gravity temperature; (d) EUP model with standard (broken lines) and modified (solid lines) black hole radius.} 
 \label{numtemp}       
\end{figure}

The modification of the HUP relation will have an impact on the various thermodynamic characteristics of black holes.  We recall that  
the Hawking temperature can be derived from the HUP 
by associating the emitted particle energy with its momentum, $E \sim c \, \Delta p$, and then determining the temperature as $T_H \sim E/k$:
\be
\Delta x \Delta p \sim \hbar ~~\Longrightarrow~~ T _H\sim \frac{\hbar c}{k\Delta x} \sim \frac{\hbar c^3}{2GMk} \, ,
\ee
where we assume $\Delta x \sim R_S = 2GM/c^2$ at the last step. 
As argued by Adler~\cite{Adl10}, the same reasoning can be used to find the Hawking temperature for the GUP:
assuming $\RS = 2GM/c^2$ and $T_{\rm GUP} \sim c \, \Delta p_{\rm GUP}/k$ gives 
\be
T_{\rm GUP} \sim \frac{Mc^2}{\alpha k}\left(1\pm \sqrt{1-\frac{\alpha M_{\rm P}^2}{2M^2}}\right) \, .
\label{AdlerT}
\ee
Choosing the negative sign gives the usual Hawking temperature in the super-Planckian regime ($M \gg M_{\rm P}$): 
\be
T_{\rm GUP} \sim \frac{Mc^2}{\alpha  k}\left(\frac{\alpha}{4}\frac{M_{\rm P}^2}{M^2}\right) \sim \frac{M_{\rm P}^2c^2}{4  M k}\sim \frac{\hbar c^3}{4GM k} \, .
\ee
Choosing the positive sign gives 
\be
T_{\rm GUP} \sim  \frac{2 Mc^2}{\alpha  k}
\label{superP}
\ee
for $M \gg M_{\rm P}$. There is no sub-Planckian regime because the square-root term becomes complex for $M < \sqrt{\alpha/2} \, M_{\rm P}$. The $T(M)$ relation is therefore as indicated qualitatively in Fig.~\ref{flatGUP}(a) and more precisely in Fig.~\ref{numtemp}(a).  Its qualitative form can be inferred by rotating  the $\Delta x (\Delta p)$ diagram in Fig.~\ref{GUP/EUP}(a)
through 90 degrees.  
Indeed, this applies for any form of the Uncertainty Principle.  If one associates $\Delta p$ with the black hole temperature and $\Delta x$ with the black hole horizon size, $2GM/c^2$, the function $\Delta x(\Delta p)$ immediately  implies the function $M(T)$ and inverting this then gives the temperature function  $T(M)$ as a 90 degree rotation.

In the BHUP approach, this derivation does not apply because one must also change the expression for $R_H$.
Associating $\Delta p$ with $kT/c$, as in the Adler approach, and $\Delta x$ with the black hole radius gives
\be
\Delta x \sim \frac{ \alpha GM}{2c^2} + \frac{\hbar}{2Mc} \sim \frac{\hbar c}{kT} +\frac{\alpha G k T}{2c^4} \, .
\ee
This is a consequence of the BHUP requirement that the same parameter $\alpha$ appears in both the black hole radius and GUP expression. This implies the {\it exact} relationship
\be
 T_{\rm BHUP} =
\begin{cases}
  Mc^2/k & (M < M_{\rm P} /\sqrt{ 2 \alpha}) \\
 \hbar c^3/(2 G k \alpha M)  & (M >  M_{\rm P} /\sqrt{2 \alpha}) \, .
\label{BHUPtemp}
\end{cases}      
\ee
and this is shown qualitatively in Fig.~\ref{flatGUP}(b) and more precisely in Fig.~\ref{numtemp}(b).  
For a more general form of the Uncertainty Principle,  
the function $\Delta x(\Delta p)$ still implies both the function $R_{H}(T)$ and $R_{H}(M)$, so it remains true that these functions must have the same form.  Clearly $T \propto M$ remains a possible solutions but $T \propto M^{-1}$ seems to be specific to the BHUP case.

On the other hand, there is another way of deriving the temperature in the BHUP case and this gives a slightly different result.   Associating the temperature with the surface gravity of the black hole implies
\be
T_{\rm BHUP} \sim (M + \beta' M_{\rm P}^2/2M)^{-1} \propto 
\begin{cases}
1/M  &  (M \gg M_P) \\
 M/ \beta' & (M \ll M_P) \, ,
 \end{cases}
\label{CMNtemp}
\ee
where $\beta'$ is used to distinguish it from the EUP parameter.  This is the expression used in Ref.~\cite{CMN15} and is shown in Fig.~\ref{numtemp}(c).  It gives the same asymptotic form as Eq.~\eqref{BHUPtemp} but is slightly different at the peak. 
Both forms are represented qualitatively in Fig.~\ref{flatGUP}(b), which can
can be inferred by turning the $R_H(M)$ diagram through 180 degrees.  

For the EUP,  one requires $M > \sqrt{\beta} M_{\rm P}$ and Eq.~\eqref{EUP}, together with the assumption that $\Delta p \sim  kT/c$, gives a Hawking temperature
\be
T_{\rm EUP} \sim \frac{\hbar c}{k R_H} + \frac{\beta c^4 R_H}{G k} \, .
\label{EUPtemp1}
\ee
For the standard Schwarzschild expression for the black hole radius, this gives
\be
T_{\rm EUP} \sim \frac{ \hbar c^3}{G k M} + \frac{\beta M c^2}{k} \, ,
\label{EUPtemp}
\ee
This is shown qualitatively in Fig.~\ref{flatGUP}(c) and more precisely by the broken lines in Fig.~\ref{numtemp}(d).  
The temperature has a minimum of $ \sqrt{\beta} \, T_{P}$ at a mass $\Mpl/ \sqrt{\beta}$ and, as anticipated above,  the qualitative form can be obtained by turning the $\Delta x(\Delta p)$ curve in Fig.~\ref{GUP/EUP}(b) through 90 degrees.  However,  if we use the modified black hole radius, given by Eq.~\eqref{EUPradius},  the functions $R_H(M)$ and $R_H(T)$ must be identical and this implies
\be
k T \pm \sqrt{k^2 T^2 - \beta \Mpl^2 c^4} = M c^2  \pm \sqrt{M^2 c^4 - \beta \Mpl^2 c^4} \, .
\ee
If one chooses the same sign on both sides,  there is clearly a solution 
\be
T_{\rm EUP} =  Mc^2/k \quad (M > \beta^{1/2} M_{\rm P}) \, .
\label{EUPtemp2}
\ee
This is shown qualitatively in Fig.~\ref{flatGUP}(d) and more precisely by the solid line  in Fig.~\ref{numtemp}(e).  However,  unlike in the BHUP case,  there is no $M^{-1}$  solution if one chooses different signs.  Also Eq.~\eqref{EUPtemp2} has no dependence on $\beta$ except in so much as the endpoint depends on $\beta$,  the minimum mass being $\sqrt{\beta} M_{\rm P}$.  By contrast,  Eq.~\eqref{EUPtemp} has a a minimum at a mass of $M_{\rm P}/\sqrt{\beta}$,  which is much larger for $\beta \ll 1$.  

\begin{figure}[h]
(a) \includegraphics[scale=.35]{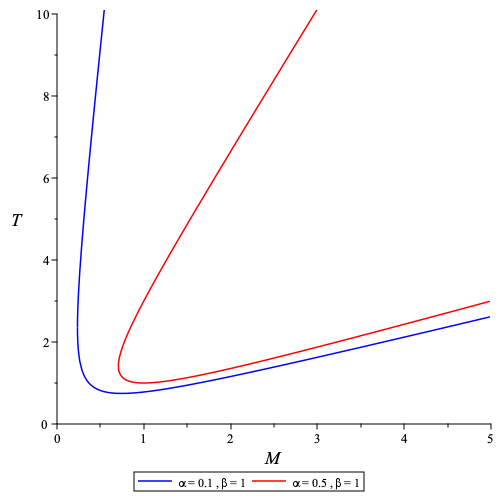}~~~
(b) \includegraphics[scale=.35]{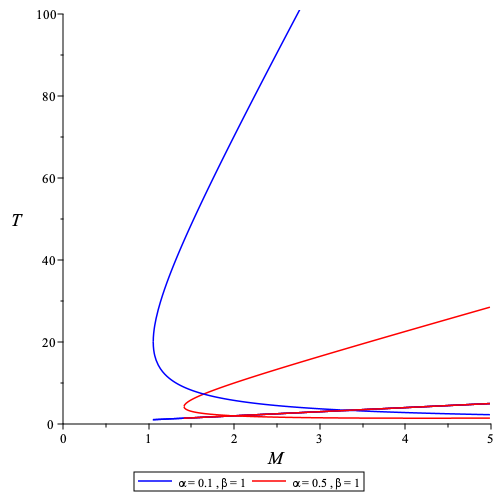}\\
(c) \includegraphics[scale=.35]{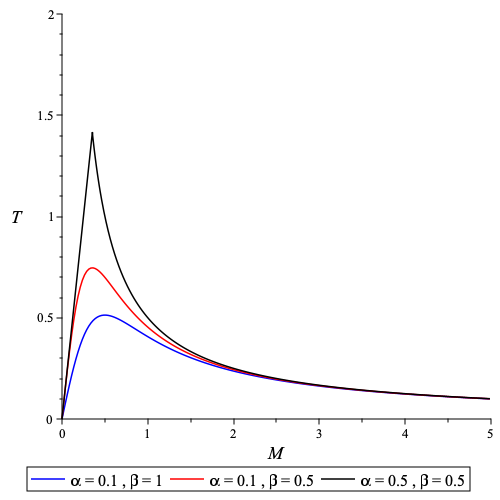}~~~
(d) \includegraphics[scale=.17]{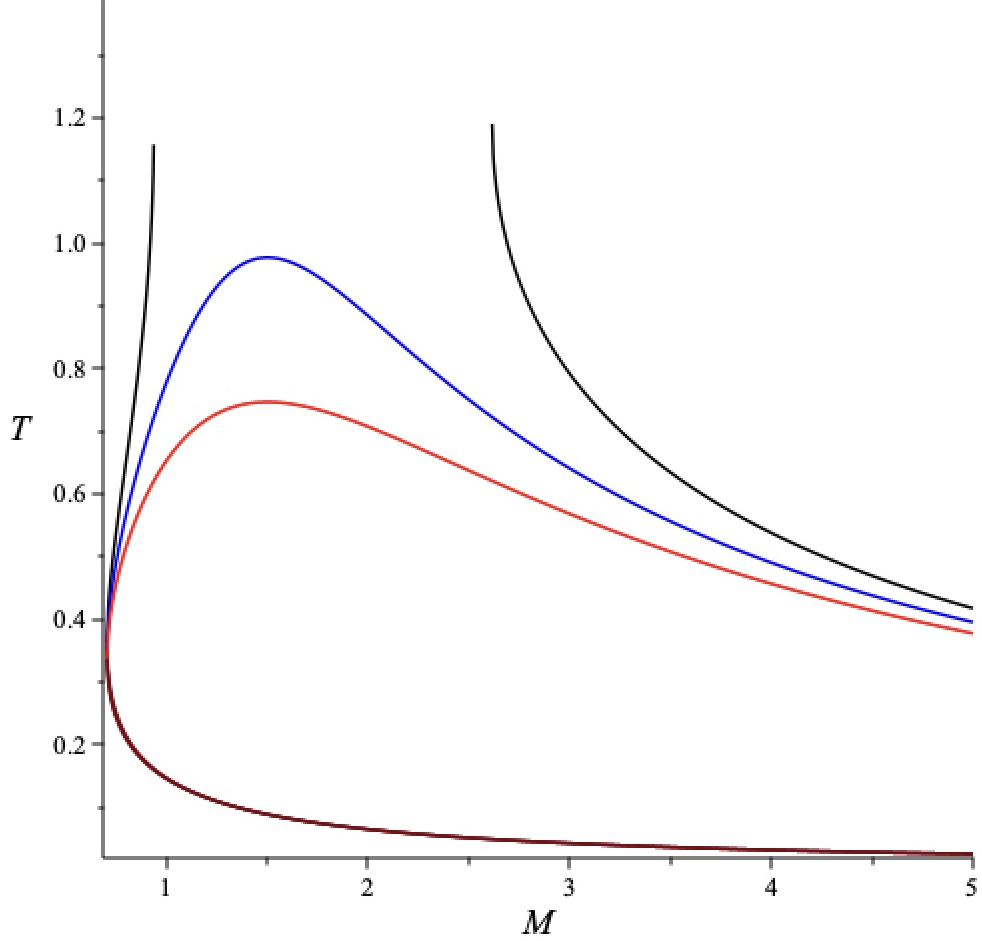}
\caption{Numerical calculations of the temperature function $T(M)$ for: 
(a) GEUP model with standard black hole radius; (b) GEUP model with modified black hole radius; 
(c) EGUP model with standard black hole radius; (d) EGUP model with modified black hole radius for
$\alpha = 0.1, \beta = 0.5$ (red line), $\alpha = 0.4, \beta = 0.5$ (blue line) and $\alpha = 0.7, \beta = 0.5$ (black line).}
 \label{numtemp2}       
\end{figure}

We now derive the
temperature expression for the GEUP relation.  
If one uses the Adler approach, with the black hole radius having the usual Schwarzschild form, then Eq.~\eqref{AdlerT} is replaced by
\be
T_{\rm GEUP} = \frac{M c^2}{\alpha k} \left(1 \pm \sqrt{1- \alpha \beta - \frac{\alpha \Mpl^2}{M^2}} \, \right) \, .
\label{GEUPtemp}
\ee
There is a minimum mass of $\sqrt{\alpha/(1- \alpha \beta)} \, \Mpl \approx \sqrt{\alpha} \, \Mpl$ and the form of $T(M)$ can be deduced directly from Fig.~\ref{GEUP} since $\Delta x \rightarrow M$ and $\Delta p \rightarrow T$. This is shown in Fig.~\ref{numtemp2}(a).
 If the black hole radius is specified by Eq.~\eqref{GEUPrad},  then the temperature is given implicitly  by 
\be
T_{\rm GEUP} = \frac{c^4 \Delta x}{\alpha k G} \left(1 \pm \sqrt{1- \alpha \beta - \frac{\alpha \ellp^2}{\Delta x^2}} \, \right) 
\ee
\be
\Delta x = \frac{G M}{\beta c^2} \left( 1 \pm \sqrt{1- \alpha \beta - \frac{\beta \Mpl^2}{M^2}} \right) \, .
\ee 
This gives Fig.~\ref{numtemp2}(b).  
 
Lastly, we obtain a temperature expression for the EGUP case of Section~\ref{Sec:EGUP}. Proceeding as before,
we find the Hawking temperature
\be
T_{\rm EGUP} \sim  \frac{c^2}{\alpha k}\left(M + \frac{\beta M_P^2}{4M}\right)\left(1-\sqrt{1-\frac{\alpha M_P^2}{\left(M + \frac{\beta M_P^2}{4M}\right)^2}}\right) \, 
\label{EGUPtemp}
\ee
if the black hole has the standard radius.  
When $M \gg \sqrt{\beta} M_P$, this becomes
\be
T_{\rm EGUP} \sim \frac{M c^2}{\alpha k} \left(1-\sqrt{1-\frac{\alpha M_P^2}{M^2}}\right) \sim \frac{c^2 M_P^2}{2 k M} \, ,
\ee
which is the standard Hawking form.  In the sub-Planckian limit,  the square-root term in Eq.~\eqref{EGUPtemp} is real for $M < \beta M_P/(4 \sqrt{\alpha})$ and in this regime the temperature becomes
\be
T_{\rm EGUP} \sim \frac{2\hbar c M}{\beta k \ellp^2M_P^2} \sim \frac{Mc^3}{k \beta} \, .
\ee
This is shown in Fig~\ref{numtemp2}(c).
For a modified black hole radius,  the factor $M$ in Eq.~\eqref{EGUPtemp} is replaced by $R_{\rm H}$, which is itself given by Eq.~\eqref{EGUPradius}.  The temperature is then as shown in Fig~\ref{numtemp2}(d) for fixed $\beta$ but varying $\alpha$; a similar range of curves is found for for fixed $\alpha$ but varying $\beta$, with a discontinuity in the behavior for $\alpha = \beta$.

\section{Conclusions}
In this paper, we have considered various extensions of the 
Uncertainty Principle: the well-known GUP and EUP; the less-studied combination of these, termed the GEUP; and a new combination termed the EGUP.  These extensions are mainly motivated by an attempt to link black holes with the Uncertainty Principle, in what is generically termed the BHUP correspondence. This works well for the GUP, where the Compton wavelength for sub-Planckian masses naturally merges with the Schwarzschild radius for super-Planckian masses.  It does not work for the EUP, since either the usual black hole radius is much larger than Schwarzschild value or the Compton wavelength is only defined in the super-Planckian regime.  However, the correspondence is restored for the GEUP and there are then three black hole phases:
the standard Schwarzschild phase;
the sub-Planckian Compton phase;
and a third phase corresponding to black holes in a strong gravity regime.  We identify the latter as a new class of (quantum) black holes. In the EGUP case, there are two distinct regimes that describe sub-Planck-mass and super-Planck-mass black holes, though the former may be unphysical. 

Finally, we should raise the more general issue of whether it really makes sense to think of every line in the $R_{\rm H}(M)$ or $\Delta x(\Delta p)$ diagram as a black hole.  If not, the meaning of the associated $T(M)$ curves is also unclear.  In Fig. ~\ref{GUP/EUP}(a), one can regard the left line as a sub-Planckian black hole because there is a specific metric, given by Eq.~\eqref{gupmetric}, to justify this.  However, we have not derived a metric for the other cases  but just assumed that the $\Delta x(\Delta p)$ relation specifies an $R_{\rm H}(M)$ relationship. For example,  in the EUP case, can the lower line in  Fig. ~\ref{GUP/EUP}(b) be regarded as a black hole?  One clearly needs to specify a metric equivalent to Eq.~\eqref{gupmetric},  which amounts to specifying a revised effective mass, equivalent to the ADM mass in the GUP case.  Conceivably, any model in which there are two distinct black hole states for the same value of $M$ is problematic.  

\section*{Acknowledgements}

JM thanks Queen Mary University of London for its generous hospitality, during the research phase and initial writing of this manuscript.

\section*{Appendix}

In the corpuscular model of Dvali and  Gomez~\cite{Dvali:2011aa}, a black hole is regarded as a Bose-Einstein condensate of N loosely coupled gravitons.  
Since no graviton can be localized within the usual Schwarzschild radius,  $\Delta x \sim R_{\rm S}$, the EUP becomes
\be
\Delta p \sim \frac{\hbar}{R_{\rm S}}\left(1 +  \beta \frac{R_{\rm S}^2}{\ellp^2} \right) \sim \frac{c M_{\rm P}^2}{M} \left(1 +  \frac{R_{\rm S}^2}{L_*^2} \right) \, .
\ee
The number of gravitons can be derived from holography as
\be
N \sim (M/M_{\rm P})^2 \, ,
\label{dvaliN}
\ee
so identifying the sum of graviton momenta with the black hole ADM mass gives 
\be
R_{H} = \frac{2 G M_{\rm ADM}}{c^3}  = \frac{2 G N \Delta p}{c^2} \approx R_{S} \left(1 + N \frac{\ellp^2}{L_*^2} \right) \, .
\ee
The second term in parentheses then gives an $M^3$ dependency \cite{Mureika_2019}, although this is not a feature of the version of the EUP considered in the main text. 

We stress that Dvali and Gomez make no mention of either the GUP or the EUP.  Their quantum N portrait (QNP) model is just a description of a black hole as a collection of gravitons and
from that they derive all the usual black hole thermodynamics. 
Indeed, one can also use the QNP approach to obtain the GUP in the form
\be
R_{H} = \frac{2 G M_{\rm ADM}}{c^2} \approx \frac{2 G M}{c^2} \left(1 + \frac{\beta'}{2N} \right) \, ,
\ee
using the relation (\ref{dvaliN})  \cite{Frassino:2016oom}.  So does QNP imply the GUP or EUP?  
The GUP correction goes as $1/N$ because this is a quantum-scale effect and so kicks in when $N$ is small. The EUP, being a large-scale effect,  turns on for large $N$.   
This applies when considering ``classicalization'', which is equivalent to the limit $\hbar \rightarrow 0$. But there is no reason why $N$ cannot be small and 
this allows the QNP to shed light on the end of black hole evaporation.

In fact, not only the GUP and EUP expressions but also their GEUP combination can be derived from the QNP. The thermal emission process for a black hole assumes two key features: the wavelength of the escaping graviton is $\lambda_{\rm esc} = \sqrt{N}\ellp$ and its energy is $E_{\rm esc} = \hbar/(\sqrt{N}\ellp)$. One can thus write the graviton's uncertainty in position and momentum as 
\be
\Delta x \sim \frac{\ellp}{\sqrt{N}}\Delta N~~~,~~~\Delta p \sim \frac{\hbar}{\ellp N^{3/2}}\Delta N \, .
\ee
The uncertainty relation is then
\be
\Delta x \Delta p \sim \hbar
\left(\frac{\Delta N}{N}\right)^2 \, .
\ee
If we assume this is a new gravitational contribution to the uncertainty relation, then adding in the Heisenberg (quantum) uncertainty gives
\be
\Delta x \Delta p \sim \frac{\hbar}{2} \left[1+ \xi \left(\frac{\Delta N}{N}\right)^2\right]
\label{qnpup}
\ee
for some parameter $\xi$ that could be related to either the GUP or EUP parameter. \\
This can be re-expressed in terms of
either $\Delta p$ or $\Delta x$, giving
\be
\Delta x \Delta p \approx \frac{\hbar}{2} \left(1+\frac{\xi\ellp^2M^2}{\hbar^2M_P^2}\Delta p^2\right)
\ee
for the GUP
and 
\be
\Delta x \Delta p \approx  \frac{\hbar}{2} \left(1+\frac{\xi}{R_S^2}\Delta x^2\right) \, ,
\ee
for the EUP.
Combining these would give a GEUP-like equation.

Reference~\cite{Pantig:2024asu} considered a form of the EUP linked to general relativity known as the Asymptotic Extended Uncertainty Principle (AEUP), whose uncertainty depends on curvature invariants of the spacetime:
\beq
\Delta x \Delta p \geq \pi\hbar\sqrt{1-\frac{R}{6\pi^2}-\xi\frac{\cal C}{\pi^2}\Delta x^4}
\label{aeup}
\eeq
 where $R$ is the Ricci scalar, ${\cal C}$ is the Cartan invariant and $\xi$ is a dimensionless constant.
The EUP itself has also been shown to give rise to the R\'enyi entropy and temperature \cite{Moradpour_2019,Nakarachinda:2022gsb}.

\bibliographystyle{apsrev4-1}
\bibliography{myrefs}

\end{document}